\documentclass[journal=jacsat,manuscript=article]{achemso}

\usepackage[version=3]{mhchem} 
\usepackage{xcolor}
\usepackage{soul}

\title{High Pressure Suppression of Plasticity due to Over-Nucleation of Shear Strain}
\author{Brenden W. Hamilton}
\affiliation{Theoretical Division, Los Alamos National Laboratory, Los Alamos, New Mexico 87545, USA}
\email{brenden@lanl.gov}
\author{Timothy C. Germann}
\date{October 2022}

\begin{document}

\maketitle

\begin{singlespace}
\begin{abstract}

High pressure shear band formation is a critical phenomenon in energetic materials due to its influence
on both mechanical strength and mechanochemical activation.
While shear banding is know to occur in a variety of these materials, the governing dynamics of the mechanisms is not well defined for molecular crystals.
We conduct molecular dynamics simulations of shock wave induced shear band formation in the energetic material 1,3,5-trinitroperhydro-1,3,5-triazine (RDX) to assess shear band nucleation processes.
We find, that at high pressures, the initial formation sites for shear bands 'over-nucleate' and rapidly lower deviatoric stresses prior
to shear band formation and growth. This results in the suppression of plastic deformation. A local cluster analysis is used to quantify and contrast
this mechanism with a more typical shear banding seen at lower pressures.
These results demonstrate a mechanism that is reversible in nature and that supersedes shear band formation at increased pressures.
We anticipate that these results will have a broad impact on the modeling and development of high strain rate application materials
such as those for high explosives and hypersonic systems.

\end{abstract}
\newpage

\section{Introduction}

Shear bands, regions of localized plasticity induced from high shear stresses, form to accommodate further deformation in materials and can lead to the buildup of material damage, as well as a potential decrease in flow stress\cite{yan2021shear}.
Shear banding has been known to occur in a variety of materials classes, including crystalline metals\cite{hu2023amorphous,wei2002evolution,jonnalagadda2010strain}, ceramics\cite{reddy2013atomic,chen2003shock,hsiao2019shear}, 
metallic glasses\cite{moorcroft2011age,greer2013shear,lewandowski2006temperature,bei2006softening,pan2020strain,wang2018spatial,ding2017computational}, and molecular materials\cite{Cawkwell2008ShearBand,wu2008numerical,Cao2012Shear}.

Overall, shear banding generally occurs when there is a local shear induced softening of the material, creating regions that are much more susceptible to a rapid material flow\cite{greer2013shear}.
This leads to a localized, plastic instability that lowers the deviatoric stress components of the material and allows for continued deformation.
Shear banding has been of more recent interest due to its relevance in metallic glasses, as well as the observance of shear band formation without prior defect accumulation\cite{luo2019plasticity}.
In the latter case, these shear bands were found to allow for continued deformation in the absence of dislocations without inducing cavitation or material failure.

A recent study has shown that shear bands nucleate to accommodate deformation (without defect accumulation) when the material has a low difference in energy between the crystal and
amorphous phases, such that forming amorphous bands is energetically favorable to other mechanisms\cite{hu2023amorphous}.
Shear banding sans crystalline defects is also well studied in polymers, in which glassy and entangled molecules can flow, even below the glass transition state\cite{Cao2012Shear,wu1994analysis}.
Molecular solids, which are often utilized at ultra-high strain rates, pose a grand challenge in studying shear band formation due to their complex crystal packing and limited crystalline defect formation mechanisms.

In general, for energetic materials, shear bands can induce a mechanochemical acceleration of kinetics\cite{Kroonblawd2020ShearBands,Hamilton2021Review,Manaa2003ShearInduced},
making shear banding a highly relevant and complex chemical initiation mechanism.
Mechanochemistry in energetic molecular solids typically occurs via a shock wave creating 
large intra-molecular strains via local material flow\cite{Gilman1993ShearInduced,Gilman1995Chemical,Hamilton2021HotspotsBetterHalf,Hamilton2022PEHotspot,hamilton2023intergranular}.
These strains then lead to a lowered activation barrier and accelerated kinetics in areas such as pore collapses and shear bands\cite{Wood2015UltrafastChemistry,frey1982initiation,Hamilton2022Extemporaneous,Hamilton2022ManyBody}, 
and they can occur in other materials such as graphite\cite{Kroonblawd2018MechanochemicalGraphite,hamilton2023interplay} and mechanophores\cite{Davis2009ForceInduced,hamilton2022rapid}.

Here, we study shear banding in a molecular crystal, specifically the energetic material 1,3,5-trinitroperhydro-1,3,5-triazine (RDX).
Shear banding is known to occur in RDX\cite{Cawkwell2008ShearBand}, and occurs due to local thermal softening, not defect buildup\cite{izvekov2021bottom}.
These shear bands form out of regions that initially show a rise in molecular conformation changes that leads to local softening\cite{izvekov2022microscopic},
and the associated free energy changes are indicative of a local first-order structural phase transition\cite{izvekov2022microscopic}.
Here, we show an additional mechanism in which higher pressure shear bands are suppressed due to an over-formation of shear band nucleation sites that
significantly lowers the deviatoric stress state. This rapidly reduces the driving force for shear band growth and no significant plasticity occurs at higher pressures,
in contrast to the significant plasticity seen at lower shock pressures.

\section{Results and Discussion}

We conduct shock compression simulations using all-atom molecular dynamics, compressing RDX along the [100] crystal direction for particle velocities ($U_p$) ranging from 0.7 to 1.2 km/s.
Figure 1 shows molecular (center of mass) renderings of these simulations, colored by the molecular shear strain (von Mises invariant of the Green-Lagrange strain tensor), at both the time of maximum compression when the shockwave reaches the far surface,
and 50ps later, with the system held at maximum compression using shock absorbing boundary conditions (SABCs)\cite{Cawkwell2008ShearBand,Zhao2006Atomistic,Hamilton2021HotspotsBetterHalf}.

These localizations of shear strain, in all cases, form a band-like structure. For the weaker shock strengths, this manifests in a few thin bands of high shear strain, 
illustrated by the orange and red regions in Figure 1.
Conversely, in the stronger shock cases, where resolved shear stresses are heightened, the shear strain localizations strikingly lack intensity, but are also considerably more pervasive.
Supplemental Materials section SM-3 presents a 1D binning (along the shock direction) of the shear strain, displaying that the stronger shock cases result in a greater average shear strain per molecule (or volume),
despite a lack of appreciable shear banding. In the stronger shock cases, running the simulation for significantly longer times does not result in more localized shear or any shear band formation (see Supplemental Materials section SM-2),
such that, by 50ps after maximum compression, the system strain is predominately static, regarding any continual structural evolution.

\begin{figure}[htpb]
  \includegraphics[width=0.5\textwidth]{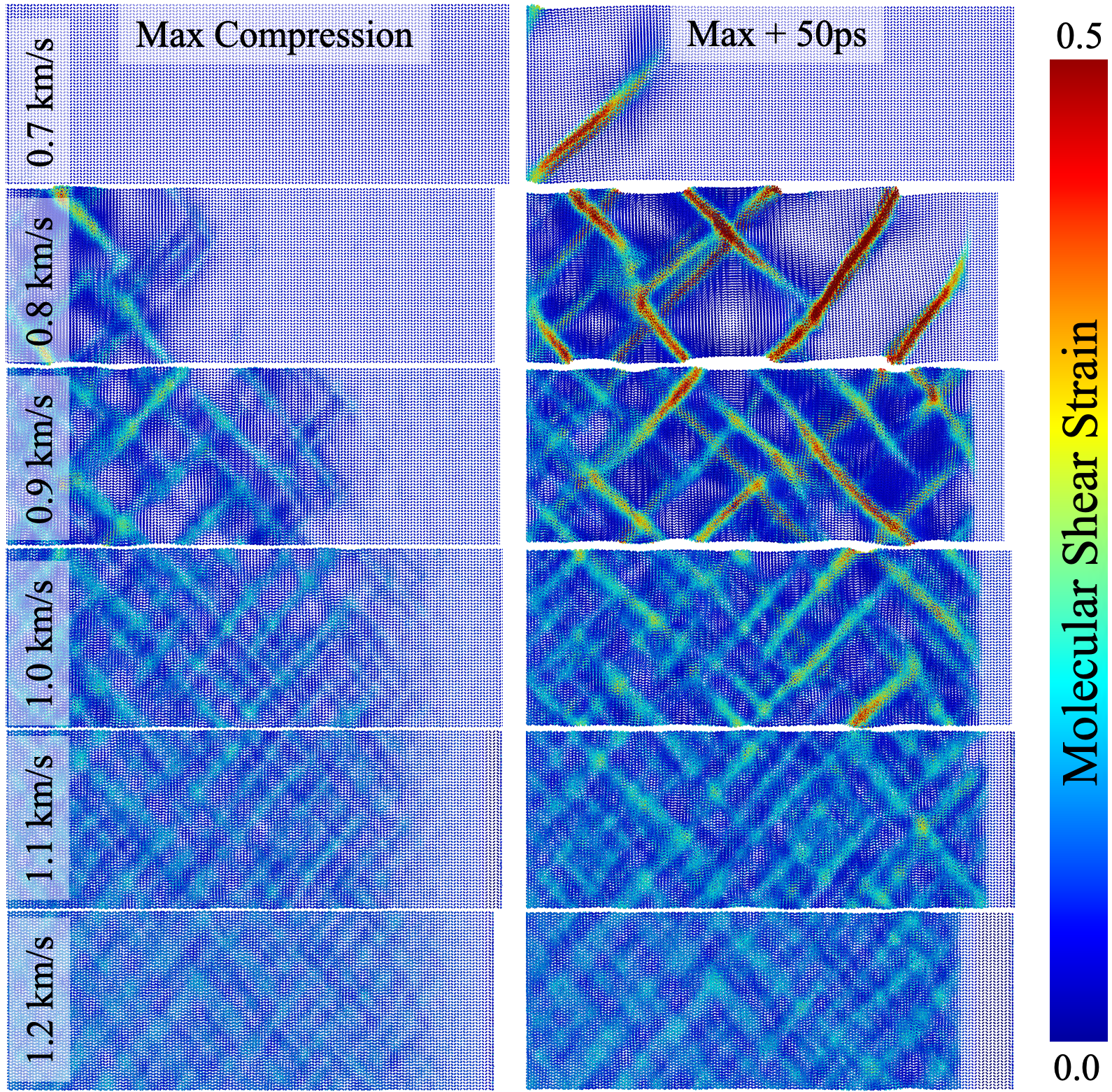}
  \caption{Molecular renderings of the X-Z plane for the time of maximum compression (left) and 50ps later, held at that compression (right),
  molecules are colored by the shear strain (von Mises invariant of the Green-Lagrange strain tensor). A figure of the same frames with a narrower color bar range is available in Supplemental Materials section SM-2.}
  \label{fig:Fg1}
\end{figure}

Hence, here we observe conditions in which the lower pressure systems exhibit significant, localized plasticity, however, increases in the uniaxial compression rate result in little to no plasticity, just a nominal shear deformation
to a significant number of molecules, suppressing plasticity with increased pressure. Additionally, for the weaker shocks, the shear banding occurs significantly behind the shock front, whereas the more homogeneous deformation of the
stronger shock cases occurs nearly adjacent to the rise in pressure from the shock. Furthermore, these stronger shock cases result in the retainment of the crystal structure, not a bulk plasticity event.

Figure 2 shows a time history of the von Mises (VM) stress for a planar slice of material 2.5nm wide, beginning at 10.0nm from the piston.
For 0.7 km/s, which does not shear band in the time of the shock passage of the entire sample, which is ~24 ps for this case, the VM stress does not significantly change.
Shifting to the prompt shear banding cases of 0.8 and 0.9 km/s, the VM stress sits at its peak value for ~5-10ps, before slowly decaying as plasticity begins to materialize, resulting in a much more hydrostatic state during the maximum compression hold.

For the strongest shock cases, a rapid decrease in VM stress is observed, occurring on the timescale of the shock rise, decreasing so promptly that the peak value shown lowers with rising shock strength.
However this is an artificial smoothing due to the bin size and temporal spacing of the points, where material in the upstream side of the bin commences decaying its VM stress prior to total compression of the bin.
This rapid decrease occurs, sans plasticity of any meaningful quantity, and the driving force for shear band formation, large deviatoric stresses, fails to be sustained in a requisite intensity to induce shear band formation.
Hence, this deformation mechanism that does occur usurps the shear banding mechanism and nullifies the driving force before significant plasticity can occur.

Rapid VM stress decays like that show here for strong shocks may also indicate a phase transformation.
Based on experimental phase diagrams\cite{dreger2010phase} for RDX, all the states studied here should exist in the $\gamma$ phase (the highest shock temperature reached here is ~420 K).
Supplemental Materials section SM-4 shows radial distribution function g(r) curves for the initial state and all shocked states, further showing that no phase transition occurs at the stronger shocks, relative to the weaker shocks.
Supplemental Materials section SM-4 also presents molecular maps of molecular rotations, showing that no other explicit phase change mechanism is occurring.

\begin{figure}[htpb]
  \includegraphics[width=0.5\textwidth]{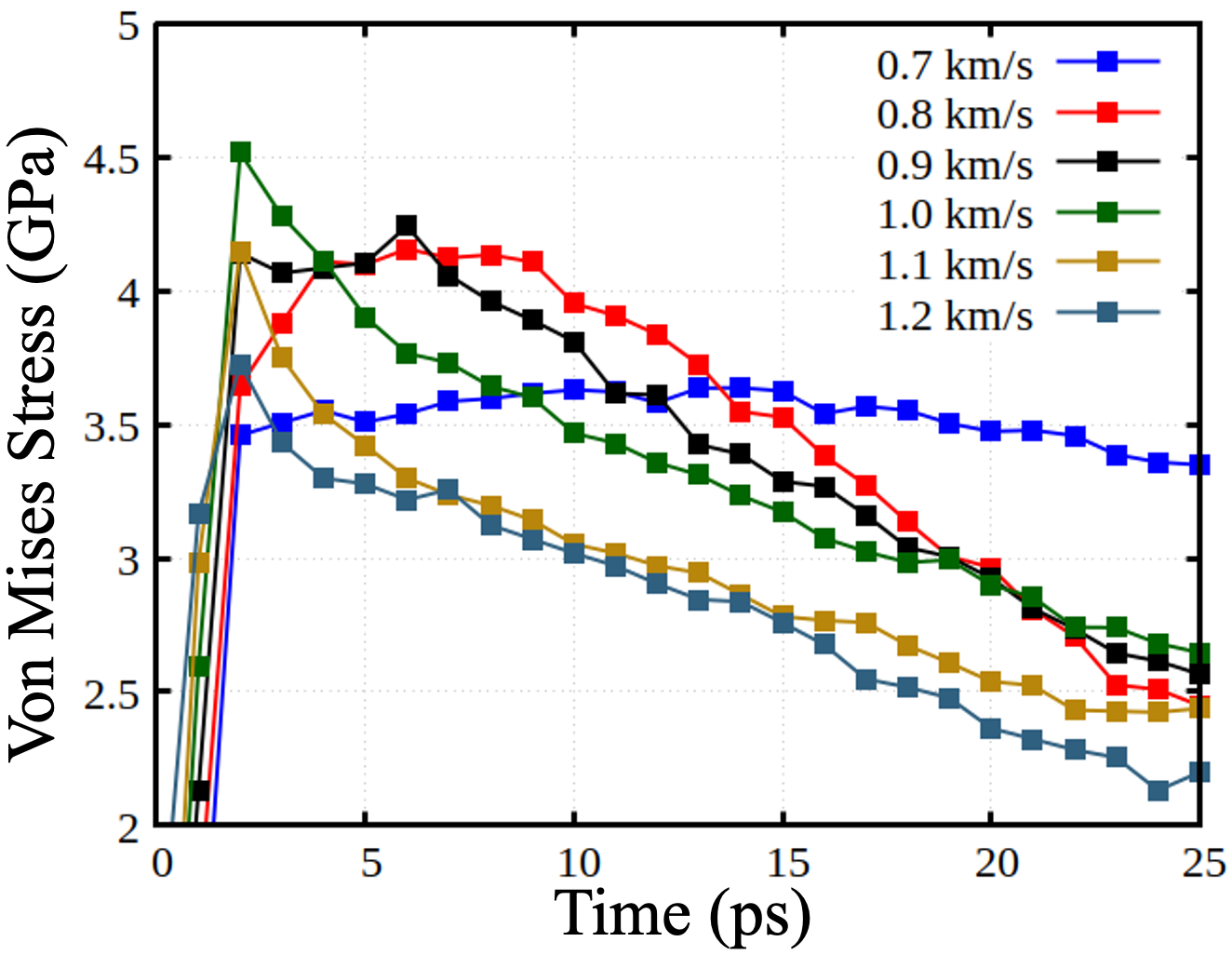}
  \caption{Time history of the von Mises stress of a 1D slice of material at 10.0 to 12.5 nm from the piston for all particle velocities.}
  \label{fig:Fg2}
\end{figure}

Figure 3 shows molecular renderings of a release of the SABCs, permitting the material to expand back towards ambient density, for a shear band and high pressure no shear band case, 0.8 and 1.1 km/s, respectively.
The SABC on the downstream face of the material is lifted after 50ps of continuous compression, and the material begins to uniaxially expand in a rarefaction process.
In the 0.8 km/s case, the shear bands neither recover to a crystalline phase nor initiate failure via void formation/cavitation.
The shear bands persist throughout the total release of the material (bottom row, +80ps) and continue to exhibit some strength and binding of the material.

Conversely, with the 1.1 km/s case, the regions of deformation, which can be discerned as the regions of increased color/darkness due to obfuscation of the white background,
present an entirely alternative release response.
As the release wave alleviates the latent stress of the compressed system, a substantial majority of these local deformations undergo an annealing-like processes and reverts to the original crystal structure.
Supplemental Materials section SM-5 shows the decrease rate in sheared molecules for both cases in Figure 3.
These results further propound the idea of plasticity suppression at higher shock strengths, such that the release wave rapidly eliminates the presence of sheared molecules in the 1.1 km/s case.

Overall, this reveals the high pressure mechanism to be reversible in nature, and, therefore, by definition, non-plastic.
This non-plastic shear deformation occurs at high enough density and has broad enough spatial tendencies to drive down the local deviatoric stresses enough to suppress plasticity, preventing the nucleation site from overcoming the high barrier for shear band formation and growth.
We spend the rest of this work attempting to characterize this high pressure mechanism.

\begin{figure}[htpb]
  \includegraphics[width=0.9\textwidth]{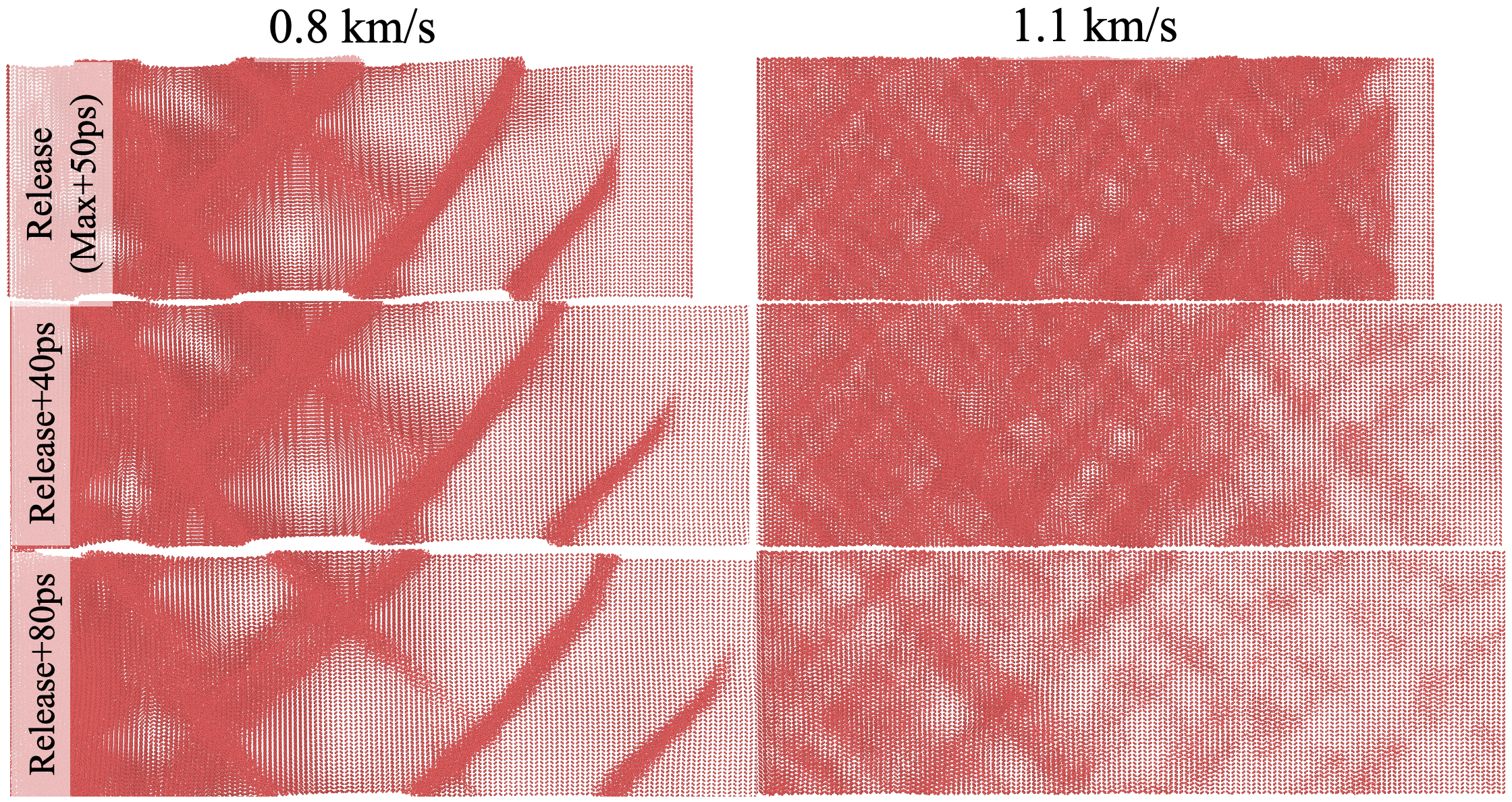}
  \caption{Molecular center of mass renderings for the release of the compressed systems, starting from 50ps after maximum compression, until full release, for the 0.8 and 1.1 km/s cases.}
  \label{fig:Fg3}
\end{figure}

To arbitrarily classify and study only the highly sheared regions of the material, we remove all molecules that have a shear strain of less than 0.15.
Figure 4 displays molecular renderings of this reduced system at 50ps after maximum compression.
The weaker shocks (top row) develop clear bands of amorphous material that span the width of the sample and consist of both 'upwards' and 'downwards' propagating shear bands.
These bands interweave with one another to form a shear band network.
By 1.0 km/s, these bands begin to increase in density sufficiently such that they begin to coalesce, resulting in a less defined network structure.

In the 1.1 and 1.2 km/s cases, no well-defined structure of the sheared material exists, and the magnitude of the shear strain is considerably lower on average.
For these cases, sheared material appears rapidly at the shock front, en masse, but never forms the defined bands of the weaker shock cases.
This is due to an 'over-nucleation' of shear band sites that never grow into plastic deformation.
Small amounts of molecules in various locations begin to shear, just as occurs for the shear bands that nucleate
in the weaker shocks. However, the number of these 'nucleation sites' is far greater and they occur much closer to the shock front.
This rapidly lowers the VM stress (as shown in Figure 2) and reduces the driving force for the shear band formation, preventing the nucleation sites from growing into shear bands, suppressing the formation of plasticity.

Supplemental Materials section SM-6 has heat map plots of the XZ Component of the Deformation Gradient Tensor (Green-Lagrange) vs the Shear Strain, in which the shear band cases show significant structure,
but the suppressed plasticity cases show no differentiation between the sheared molecules and the rest of the system.
While understanding the states in which plasticity occurs is paramount for understanding continued deformation and failure of materials in general, for energetic materials such as
RDX, there is the further pertinence in that shear bands tend to generate a mechanochemical effect that can greatly alter the material's overall reactivity\cite{Kroonblawd2020ShearBands,Hamilton2021HotspotsBetterHalf}.
The high pressure lack of plastic deformation has significant ramifications on the material strength and reactivity at its intended usage conditions.

\begin{figure}[htpb]
  \includegraphics[width=0.9\textwidth]{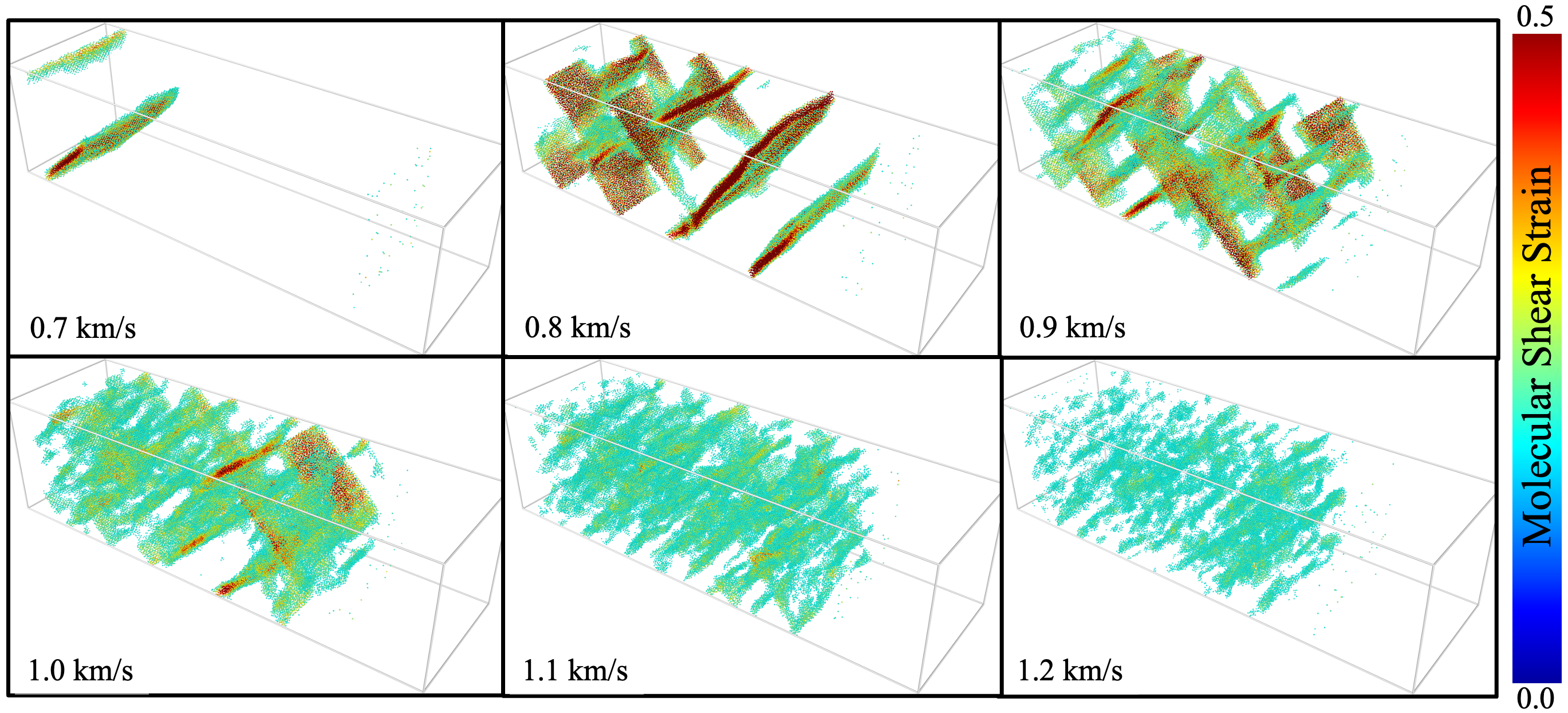}
  \caption{Molecular center of mass renderings at 50ps after maximum compression for all particle velocities. Only molecules with a shear strain above 0.15 are rendered. Molecules are colored by shear strain. }
  \label{fig:Fg4}
\end{figure}

To numerically assess the structure of the molecules with a shear strain greater than 0.15, as shown in Figure 4, we perform a cluster analysis with a distance based cutoff of 6 A.
Table 1 shows the results for each of the particle velocities with the total number of molecules considered, the number of clusters with
at least 10, 25, and 100 molecules, the size of the largest cluster, and what percentage of the considered molecules are in the largest cluster.
While we conduct this analysis here on the 50ps after maximum compression frame, we find this to also be an extremely elucidating form of analysis for early time nucleation sites of shear bands in which we find similar results.

Starting with the maximum cluster size and percent, for the cases that shear band, the largest cluster is over 90\% of all sheared atoms.
This results from the various shear bands intersecting into a network to form one large cluster. In the 0.9 km/s case, a number of small ($N>10$ molecules) clusters begin to form and do not become
part of the shear banding network. By 1.1 km/s there are several hundred small clusters that fail to form a network of sheared material or result in plasticity, with the largest cluster in the 1.2 km/s case
only accounting for 26.2\% of all sheared material.
Supplemental Materials section SM-7 shows time history plots of the max cluster size and number of clusters larger than N=25 for the initial shock. The latter shows the much larger increases
for the non-shear banding cases, as well as a significant reduction in clusters after the material begins to release, showcasing the reversible (non-plastic) nature of the mechanism.

\begin{center}
\begin{table}[]
\caption{Clustering results for a cluster analysis of the high shear strain molecules shown in Figure 4. Pop. is the total molecules used per frame. $N_X$ is number of clusters with at least X molecules.
$C_{Max}$ is maximum cluster size, and $C_{Max\%}$ is the percentage of molecules in the max cluster.}
\begin{tabular}{||c c c c c c c||} 
 \hline
 $U_p$ & Pop. & $N_{10}$ & $N_{25}$ & $N_{100}$ & $C_{Max}$ & $C_{Max\%}$ \\ [0.5ex] 
 \hline\hline
 0.7 & 16145 & 3 & 1 & 1 & 15401 & 95.3 \\ 
 \hline
 0.8 & 92627 & 7 & 1 & 1 & 88787 & 95.8 \\
 \hline
 0.9 & 109401 & 40 & 2 & 1 & 103225 & 94.3 \\
 \hline
 1.0 & 107424 & 85 & 8 & 4 & 98328 & 91.5 \\
  \hline
 1.1 & 97802 & 193 & 37 & 18 & 80091 & 81.9 \\
 \hline
 1.2 & 52902 & 372 & 102 & 65 & 13854 & 26.2 \\ [1ex] 
 \hline
\end{tabular}
\end{table}
\end{center}

To better assess when a region has been shocked, we move to a 1$nm^3$ 3D Lagrangian binning, in which each bin only contains 3-4 molecules, resulting in little smoothing of
the shear strain states. Figure 5 shows the average shear strain of each bin at 2.5 ps after the bin molecules experience the shockwave in the x-axis, with the shear strain
after 50ps held at maximum compression in the y-axis.
For the weak shocks that shear band, little shearing occurs at early times and results in the data having a highly vertical trend.
This lack of correlation is due to a majority of the shear bands growing from smaller nucleation sites.
In the homogeneous shearing cases of 1.1 and 1.2 km/s, the data follows a trend of the slope just greater than 1, showing almost no increase in shearing over time or
any rearrangement in what areas are sheared more.
The result data trends have Pearson correlation coefficients of 0.083, 0.443, 0.674, 0.803, and 0.804 for 0.8, 0.9, 1.0, 1.1, and 1.2 km/s, respectively.
Other than the initial nucleation sites, there is little to no correlation between the initial and final states of the shear banding case.
However, the shear states of the higher pressure cases shows little temporal evolution, as the non-plastic mechanism lowers the deviatoric stress components more than the shear banding,
allowing the system to remain static in this state, resulting in a high pressure suppression of plasticity.

\begin{figure}[htpb]
  \includegraphics[width=0.5\textwidth]{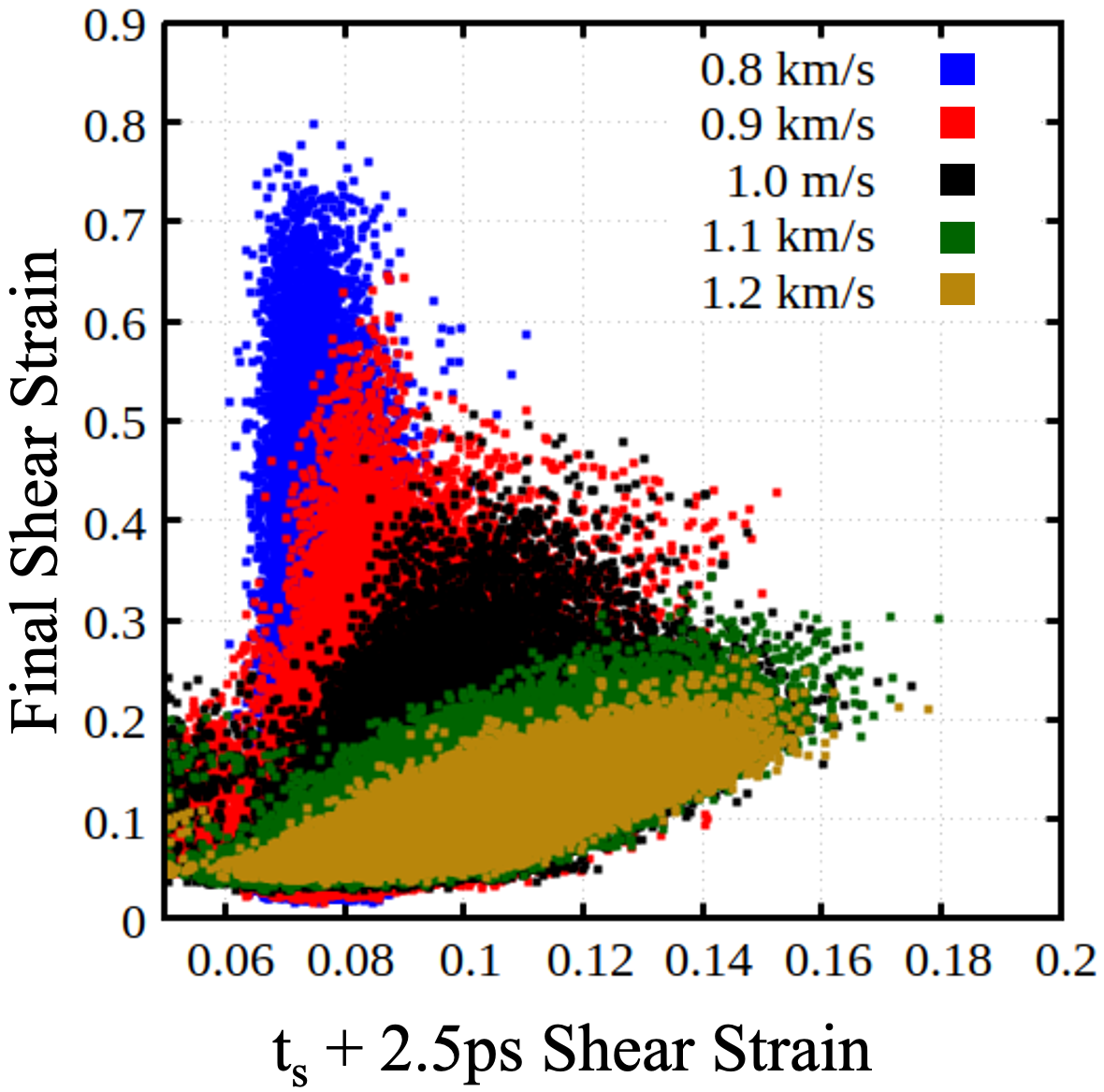}
  \caption{Correlations between the shear strain at 2.5ps after a material section is shocked vs the shear strain of that material at 50ps after maximum compression.
  Each point represents a 1$nm^3$ Lagrangian bin.}
  \label{fig:Fg5}
\end{figure}

\section{Conclusions}

Through molecular dynamics simulations of the shock compression of the energetic material RDX, we find a high pressure mechanism for the suppression of plasticity via
the over formation of shear band nucleation sites.
Upon release of the compression state, these nucleation sites heal, resulting in a reversible processes.
This highly differs from lower pressure shocks in which plastic deformation results in significant shear band growth that remains upon release.
In both cases, the deviatoric stress components are lowered, however, the non-plastic, high pressure mechanism results in a more rapid decay of the von Mises stress.
Hence, the deformation mechanism that occur at higher pressures supplants the shear banding mechanism and curtails its driving force prior to any plasticity.

A cluster analysis of sheared molecules shows the shear bands to be one large, interconnected network, whereas the high pressure system yields hundreds
of small, independent clusters that do not significantly grow in size or shear magnitude over time, yet nevertheless lower the shear stress drivers for plasticity.
Overall, the lack of high pressure shear banding is critical to RDX for both mechanical strength and mechanochemical effects on kinetics.
Hence, future work will assess the broad nature of this mechanism in RDX for a variety of thermal, microstructure, and mechanical states, as
well as non-shock high-strain rate conditions and its prevalence in a variety of materials classes where high strain rate plasticity is of great importance, such as polymer binders and materials for hypersonic applications.

\section{Methods}

We use all-atom molecular dynamics using the LAMMPS package\cite{Plimpton1995LAMMPS,Thompson2022LAMMPS}. Atomic forces and energies for RDX are calculated with the
non-reactive potential from Smith and Bharadwaj \cite{Smith1999QuantumFFHMX}. A 1.0 fs timestep was used in all simulations. Cells were constructed by replicating the
alpha-RDX crystal structure along the [100], [010] and [001] directions, aligned with the x, y, and z Cartesian axis respectively, 
to lengths of 118.2nm, 34.3nm, and 31.8nm (30x10x10 unit cells), respectively. 

5.0nm of material was removed, by whole molecules, at either end of the x-direction to break periodicity and form a free surface. The system was then equilibrated for 500ps at 300 K
using a Nos\'e-Hoover thermostat\cite{Nose1984Unified}, to allow breathing modes from the free surface creation to attenuate.
Shock simulations were conducted using the reverse ballistic approach along the [100] direction\cite{Holian1979ThreeDimensionShock}. The bottom 2.5nm, by whole molecules, was held rigid to form an infinitely massive piston.
When each shock reaches maximum compression, shock absorbing boundary conditions are applied to extend the simulation indefinitely\cite{Cawkwell2008ShearBand,Zhao2006Atomistic,Hamilton2021HotspotsBetterHalf}.
All simulation analysis was conducted on a molecular center of mass framework. Additional methods and all specific analysis details are available in Supplemental Materials section SM-1.

\section{Acknowledgments}
The authors thank Marc Cawkwell for useful discussions regarding RDX shear banding and phase transformations.
Funding for this project was provided by the Director’s Postdoctoral Fellowship program at Los Alamos National Laboratory, project LDRD 20220705PRD1. Partial funding was provided by the Advanced Simulation and Computing Physics and Engineering Models project (ASC-PEM). 
This research used resources provided by the Los Alamos National Laboratory (LANL) Institutional Computing Program. This work was supported by the U.S. Department of Energy (DOE) through LANL, which is operated by Triad National Security, LLC, 
for the National Nuclear Security Administration of the U.S. Department of Energy (Contract No. 89233218CNA000001). Approved for Unlimited Release LA-UR-21177.

\section{Author Contributions}

BWH and TCG conceived the study. BWH conducted all simulations and wrote the manuscript. TCG helped with devising analysis techniques. BWH and TCG contributed to discussions and revisions of the results and manuscript.

\section{Supplemental Material}
\includegraphics[scale=0.7,page=2]{Supplemental_Materials.pdf}
\newpage
\includegraphics[scale=0.7,page=3]{Supplemental_Materials.pdf}
\newpage
\includegraphics[scale=0.7,page=4]{Supplemental_Materials.pdf}
\newpage
\includegraphics[scale=0.7,page=5]{Supplemental_Materials.pdf}
\newpage
\includegraphics[scale=0.7,page=6]{Supplemental_Materials.pdf}
\newpage
\includegraphics[scale=0.7,page=7]{Supplemental_Materials.pdf}
\newpage
\includegraphics[scale=0.7,page=8]{Supplemental_Materials.pdf}
\newpage
\includegraphics[scale=0.7,page=9]{Supplemental_Materials.pdf}
\newpage
\includegraphics[scale=0.8,page=10]{Supplemental_Materials.pdf}
\newpage
\includegraphics[scale=0.8,page=11]{Supplemental_Materials.pdf}
\newpage
\includegraphics[scale=0.8,page=12]{Supplemental_Materials.pdf}

\bibliography{references}
\end{singlespace}
\end{document}